\documentstyle[elsart12]{elsart}

\begin{document}

\begin{frontmatter}

\title{Phonons of Metallic Vicinal Surfaces}

\author{Abdelkader Kara$^*$ and Talat S. Rahman}
\address{
Department of Physics, Cardwell Hall,
 Kansas State University, Manhattan, Kansas 66506, USA}

\address{$^*$Corresponding author: e-mail: akara@ksu.edu,
 FAX: 785 532 6806}

\date{\today}
\journal{Surface Science}

\begin{abstract}
We present an analysis of the vibrational 
 dynamics of metal vicinal surfaces 
using the embedded atom method to describe the interaction
 potential and both a real space Green's function method
 and a slab method to calculate the phonons. 
 We report two main general characteristics :  a
 global shift of the surface vibrational density
 of states resulting from a softening of the force field. The latter
 is a direct result of the reduction of coordination for the 
 different type of surface atoms; and  an appearance
 of high frequency modes above the bulk band, resulting from 
 a stiffening of the force field near the step atom. The latter
 is due to a rearrangement of the atomic positions during
 the relaxation of the surface atoms yielding a large 
 shortening of the nearest neighbor distances near the step
 atoms.

\end{abstract}
\end{frontmatter}

vicinal single crystal surfaces;surface waves, phonons\\

\section{ Introduction}

Steps  at crystal surfaces have been known to play an important role in several phenomena including
epitaxial growth and heterogeneous catalysis.
The soaring of the technological advances occurring during the last 
 few decades is due to a better understanding of the behavior of
 materials. The properties of certain materials become richer
 as their structures become more complex. The simplest aspect
 of complexity in materials structure is the loss of neighbors
 resulting in a reduction of coordination.
 A simple picture would imply that
 the lower the coordination, the larger
 the rearrangement of electronic and ionic structures.
 In the case of stepped surfaces in particular, it was pointed out by
Smoluchowsky \cite{smo} that
an electronic relaxation
at the step could lead to a de-population of surface states at that
locality, a view  supported later by several calculations
\cite{ters81} -\cite{meme}.
 The lower
coordination and the reduced symmetry of atoms on vicinal surfaces, as
compared to those in the bulk solid and on flat surfaces, lead
to characteristic variations in the surface electronic charge densities
which may in turn affect the reactivity and the propensity to harbor
localized vibrational and electronic surface excitations.  Early interest
in vicinals
was motivated by the need to comprehend in a rather controlled
manner the sensitivity of catalytic reactions \cite{somor} to atoms in step
and kink sites. Attention on vicinal
surfaces has renewed in recent years, because of their technological
importance as templates \cite{ernst1} for the growth of well ordered,
laterally patterned nanostructures.  For such technological purposes there
is also the need to understand the stability of
 vicinal surfaces as a function of
surface temperature.
The presence of
the surface, the step, and the kink can be thought of as perturbations
of increasing complexity on an otherwise periodic system. On flat surfaces
the reduction in translational symmetry in the direction
perpendicular to the surface leads to localized surface vibrational modes
whose frequencies, though distinct from the bulk modes, lie within
 the bulk phonon band.
 The frequencies of surface localized modes at Brillouin zone boundaries
 have been of particular interest as the extent to which they lie below
 the bulk band serves as a measure of the changes in surface force
 constants from values in the bulk \cite{iba91}.
  Recent experiments using
 He atom-surface scattering technique
 have also unveiled such
  low frequency, step localized mode on
 vicinals of Cu \cite{wit95} and
Ni \cite{niu95}.  On the other hand, in early efforts to
measure phonons at metal surfaces Ibach and Bruchmann \cite{iba78} had
found a mode on Pt(775), using electron energy loss
 spectroscopy (EELS), whose frequency was higher than bulk phonon band.
 It is the purpose of this paper to demonstrate that these two
 opposite behaviors (a shift towards low frequencies and a stretch
 beyond the top of the bulk band) of the vibrational dynamics are general
 characteristics of vicinal surfaces.

 The paper is divided in three sections. In the next section, we present 
 details about the geometry of vicinal surfaces and theoretical details.
 In section III, we present the vibrational dynamics of Ni(977)
 as an illustration of the softening of the force field near the
 step. Phonons of Cu vicinal surfaces will be used in section
 IV to illustrate the stiffening of the force field near the step
 and finally, in section V we present concluding remarks.

\vskip 20pt

\section{ Theoretical Details }

\vskip 12pt

Vicinals are surfaces for which the macroscopic orientation forms a small
angle with respect to a low index plane.  Ideally, they are made up of terraces
separated by regularly spaced steps of mono-atomic height. 
The Ni(977) surface is a vicinal of Ni(111)
fabricated by cutting Ni(111) at an angle of $7^o$ from the (111)
plane. It consists of (111)
terraces containing eight $<$01\=1$>$ chains, and a (100) 
step face. As shown in Fig. 1, 
the x and y axes are in the (977) plane and are, respectively, perpendicular 
to and along the step, while the z axis is perpendicular to the
(977) plane.  A side view of (977) appears in Fig 1, and the 
centered rectangular surface Brillouin zone, along with the top view, 
 is also shown in Fig. 1.

For the interaction potential between atoms in the Ni and Cu vicinals,
we use the 
many-body, embedded atom method (EAM)  \cite{daw84,daw93} potential. 
This interatomic potential
has proved to be quite successful in describing a
variety of properties in the bulk and at the surfaces and interfaces for
 six fcc metals: Cu, Ag, Ni, Pd, Pt and Au. We have also
obtained reliable results for the dynamics of 
 Cu, Ag and Ni surfaces \cite{yang}, and for
the energetics of  Cu vicinals \cite{tian93} within the EAM potentials.
 For Ni(977), as in the case of Cu vicinals, after
constructing the system in the bulk truncated positions, we apply 
standard conjugate
gradient, as well as, annealing-quenching procedures to
determine the minimum energy
configuration. The two methods yield the same minimum energy configuration,
which is found to display complex relaxation  patterns. 

 To calculate the vibrational density of states we use two methods:
 a real space Green's function method (RSGF) and a slab method,
 in the harmonic approximation. 
 The advantage
of a RSGF method is that one calculates the
total density of states with no a-priori choice of wave-vector,
 as would be the case in calculations based on 
  'k-space'.
 Also one can get 
the polarization of the vibrational
modes from the imaginary part of the columns
of the Green's functions associated with the system at hand.
 This method exploits the fact that for a system
with a finite range of interatomic interactions, the force constant matrix
can always be written in a block tridiagonal form
\cite{dy79} in which the sub-matrices along the diagonal represent
interactions between atoms within a chosen local region and the
sub-matrices along the 'off-diagonal' correspond to interactions between
neighboring localities.  Thus an infinite/semi-infinite system is converted
quite naturally into an infinite/semi-infinite set of local regions.
The real space Green's function
method also has an advantage over the familiar 'continued fraction' method
\cite{hay72} as it does not involve truncation schemes to determine the
recursion coefficients, rather a more general and simpler recursive scheme
is applied \cite{wu94}, which has been successfully used
for the study of the phonons of Au vicinals \cite{kar94}.  The vibrational
density of states $N_l(w)$ corresponding to locality "$l$" 
 is related to the trace
of the Green's function by the well known relation

\begin{equation}
\rho_l{(\omega^2)} = - {1 \over 3n_l\pi} \; \lim_{\epsilon\rightarrow0}
\left\{ Im[{\rm Tr}(G_{ll}(\omega^2+i\epsilon))] \right\}
\end{equation}
with $N_l(w)=2w\rho_l{(\omega^2)}$, 
 where G$_{ll}$ is the Green's function sub-matrix associated
with locality "$l$" and $n_l$ is the number of atoms in this locality.

   For the slab method \cite{all71} we proceed as follows. With the atoms in the
 equilibrium configuration,
 the force constant matrix
 needed to calculate the vibrational dynamics of the system is extracted
 from the partial second derivatives of the EAM potentials.
The secular equation with the dynamical matrix is then diagonalized in a
 straight forward manner to obtain
 the phonon frequencies and displacement patterns from the
eigenvalues and eigenvectors of the diagonalized matrix.

\vskip 20pt
\section{ Vibrational Dynamics of Ni(977) : softening of the force field }

\vskip 12pt

The calculated local density of states (LDOS) for the step and the terrace atoms
of Ni(977) are compared with those for the Ni(111) surface atoms in Fig. 2.
 we note that
the LDOS of the Ni(977) terrace atoms is almost identical to the one
of Ni(111). This result is a good indicator that changes due to the creation
of steps is highly localized in nature.
 The other interesting feature in Fig. 2 is a global shift of the step atom
low frequency band
to even lower frequencies, as compared to 
 the modes associated with the terrace (and 
 Ni(111) surface atoms). This is a direct signature of
 an extra softening of relevant force constants, due to a loss of
 five neighbors,
in agreement with suggestions of Niu et al. \cite{niu95}.   
 To appreciate the implications of the loss of neighbors
for the Ni(977) step atoms, we now turn to an examination of the
force constant matrices associated with them.
As seen from Table I, the reduction in the coordination
results in a large softening of the force constant matrix
elements between two step atoms 
 : 71\% along x ($k_{xx}$) and 13.7\% along y ($k_{yy}$),
 as compared to their counterparts in the bulk.
 Note that on Ni(111),  $k_{yy}$, between neighboring
 atoms in the top layer, is reduced by 5.7\%
from the bulk value.

For flat surfaces, the loss of neighbors in the direction perpendicular
 to the surface is the cause of the appearance of a surface mode
 (the so called Raleigh mode) with an atomic displacement
 perpendicular to the surface and an in plane propagation.
 This is a 
 quasi 2 dimensional mode as the displacement amplitude decays exponentially
 into the bulk. Unlike the flat surface,
 vicinal surfaces present an equivalency between atoms
 belonging to the same chain and hence,
  we expect
 that this 1D equivalency would cause the appearance
 of a quasi-1D mode.  
 The mode at 3.3 THz 
 is a quasi-one-dimensional (Q1D) with the following characteristics (Fig. 3):
it involves only the concerted motion of the step and corner atoms
 in conjunction with that of the terrace chain of atoms next to
the step (transparent atoms in Fig.3). 
 The step atoms alternate with displacement vectors 
(+1.0,0.0,-0.67) and (-1.0,0.0,+0.67),
 the corner atoms with vectors
(+0.27,0.0,-1.0) and (-0.27,0.0,+1.0), and the  terrace atoms
with vectors (0.0,+0.24,0.0) and ( 0.0,-0.24,0.0).
 Thus the step and the corner atoms move in the xz plane,
 while the terrace atoms move along the y direction to accommodate
the propagation of this mode along the y axis which lies
parallel to the step.
 The rest of the atoms are at rest.
 The wavelength of this mode is
$2\times${\em nearest neighbor distance}
 or 4.978 $\rm \AA$ and its wave-vector q= ( 0.0, 1.262 $\rm \AA^{-1}$, 0.0 ).
It is thus a mode at  $\overline{Y}$
 ( this is  $\overline{K}$ of the (111) surface Brillouin zone)
 shown in Fig. 1.
 This is a {\it super} localized mode since the motion is restricted to
the step atoms and their immediate neighbors on the surface in good agreement
 with the experimental observation \cite{niu95}.

\vskip 20pt
\section{ Vibrational Dynamics of Cu vicinals : stiffening of the force field }

\vskip 12pt

 We showed in the previous section that the step LDOS presents a global
 shift towards the low frequencies and the question is whether this global shift
 is also experienced by the high frequency band. 
 About two decades ago, Ibach and Bruchmann \cite{iba78}
 found a mode on Pt(775), using electron energy loss
 spectroscopy (EELS), whose frequency was higher than
 that of the Pt bulk phonon spectrum.
 The appearance of this mode was later shown to be due to a stiffening
 of the force field near the step \cite{all79}. 
 Such mode was also found on Au(511) \cite{kar94} and Ni(977) \cite{kar97}.
 Our recent systematic study of the vibrational dynamics and thermodynamics
 of several vicinals of the (100) and (111) surfaces of Ag, Cu, Ni
 and Pd \cite{bal01} of varying terrace width (3 to 10 atoms) provides further
 evidence that the presence of localized modes above the bulk band
 is a characteristic of vicinal surfaces of metals obeying the bond-length
 bond-order correlation. This characteristic can be traced to the stiffening
 of some force constants of the step atoms.
 In this section we present only 
 selected results for copper vicinals, of two types. The first set is that of
  Cu(511), Cu(711),
 Cu(911) and Cu(17,1,1) which are vicinals of Cu(100) with terraces
  3, 4, 5 and 9 atoms wide, respectively. The other
 set comprises  Cu(211) and Cu(331)
 which are vicinals of Cu(111) with 3 atoms-wide terraces, and,
 respectively, a (100)
 and (111)-microfaceted step face \cite{dur97}.

  To illustrate the appearance of modes above the bulk phonon band and
 its correlation to the stiffness of specific force constants, we
 consider here the case of Cu(211).
 In Fig. 4, we show the projected vibrational density of states, at the
 $\overline{\Gamma}$, $\overline{X}$ and $\overline{Y}$ points for the step (SC),
  BNN (see figure 1),
  and bulk atoms. For clarity, we present here only
 the z (perpendicular to surface) component of the Local density of states.
In Fig. 4a, at the $\overline{\Gamma}$ point, 
 near the top of the bulk band, we note two modes involving
 both the SC and BNN atoms, one mode just below
 the top of the bulk band and another one above it.
  We note that for  both modes, the amplitude is much larger
 for the BNN than for SC. 
 In Fig. 4b, at the $\overline{X}$, these high frequency modes
 present the lowest amplitude. Instead,  we find resonance
 modes inside the bulk band and localized
 surface modes beyond the bulk band and in the
 "stomach" gaps as discussed in detail previously
 \cite{kar00}.
 The largest signature of the
 high frequency mode, relative to the bulk,
 occurs at the $\overline{Y}$ point (Fig. 4c)
 in the surface
 Brillouin zone. 
 Features similar to those in Fig. 4 have been found for all vicinals
 of the two sets considered here.
 The shifts of these high frequency modes above the bulk band
 for Cu(211), Cu(511) and Cu(17,1,1) are shown in Table II.
 In the first column, we report the relative change in the bond
 length between the SC and BNN atoms and in the second column
 we show the relative change in the z-component of this bond.
 Note that in all cases the reduction of the bond length is dominated by
 a reduction of the z component of the bond.
 The shift of the frequency above the top of the bulk band
 is higher for Cu(17,1,1) than that for Cu(211) and Cu(511)
 implying a stronger relaxation of Cu(17,1,1) due to a relief
 of step-step interaction. The reason for
  choosing the three surfaces in table II was the availability of experimental data
 \cite{kar00}.
 As has been discussed in detail in Ref\cite{kar00}, EELS data reveal a mode above
 the bulk band for Cu(211) but not on the other two surfaces. We await
 further studies to settle this issue.

 To link this shift of the frequency above the top of the bulk band
 to the change of the force field near the step, we show in tables
 III and IV the force constants associated with the step atom, on the one
 hand, and the corner, the BNN and the terrace atom, on the other.
 The most important stiffening occurs for $k_{zz}$ of the force constant
 matrix between the step atom and the BNN. This is responsible for
 the large amplitude of the high frequency mode associated with the BNN
 \cite{kar00}. The stiffening of $k_{zz}$ between the SC and BNN
 in the case of (17,1,1) is twice that of Cu(211) and Cu(511) \cite{kar00}
 resulting in the larger shift mentioned above, further illustrating
 the correlation between the stiffening of this $k_{zz}$ to
 the appearance of high frequency modes. In Ref\cite{kar00},
 we also reported an even higher shift above the bulk band in the case
 of the kinked surface Cu(532) associated with an even larger stiffening
 of the appropriate $k_{zz}$.

 This large stiffening of the force field is actually due to a shortening
 of the nearest neighbor distance between the step atom and it's
 neighbors (except for the neighbor on the same step chain).
 As the step atoms loose 5 neighbors, the electronic charge density
 is strongly perturbed and a rearrangement of the electronic
 structure is accompanied by a strong rearrangement of the ionic
 positions hence lowering the total energy of the system. 
 We find the shortening of the bond length near the step to be a general
 feature of the stepped surfaces
 and independent of the terrace width and the terrace geometry.
 As an example, we illustrate in Fig. 5 the change in the bond lengths
 of the step atom with its nearest neighbors for
 vicinals with  3-atom wide terraces  with different geometries:
 Cu(511) a vicinal of (100) with a (111) step face, Cu(211)
 a vicinal of (111) with a (100) step face and Cu(331) a vicinal
 of (111) with a (111) step face. All 3 surfaces show the same
 general trend of a shortening of the bond length near the step
 \cite{dur97}. In the case of Cu(211), there is a general and quantitative
 agreement between EELS data \cite{seyller}, {\it ab initio} calculations
 \cite{wei98} and EAM calculations  \cite{dur97}. 
 In Fig. 6 we present the same effect for vicinals of (100)
 with different terrace widths. We note again that the
 bond length associated with the step atom is always shortened.

\vskip .2in

\vskip .1in

\vskip 20pt
\section{ Conclusions }
\vskip 12pt

 We have focussed in this paper on two characteristics
 of the vibrational modes of vicinal surfaces consisting of two
 distinct results: the softening  and the stiffening of the
 force constants associated with the step atoms.

 The vibrational local density of states
 (LDOS) of the step
atoms shows a distinct global shift towards the low
frequencies, as compared to that of terrace atoms, and is
 attributed to softening of some force constants
associated with the step atoms. 
 For the case of Ni(977), the result is 
 in qualitative agreement with  He scattering
data of Niu {\it et al} \cite{niu95}.

We have also presented here  arguments about the existence
and origin of modes above the bulk band on several vicinals of Cu.
Our lattice dynamical calculations show that the high frequency
modes (above the bulk phonon band)
 are a natural feature of metal vicinal
surfaces and should be present at least on those surfaces
 whose dynamics could be described adequately by potentials
 like the EAM which assume that the electronic
 charge distributions to be spherically symmetric and follow 
 bond-length bond-order correlation.
 The high frequency modes are due to a  stiffening of the force field
 near the step resulting from a global shortening of the nearest-neighbor
 distances associated with the step atom. Since the BNN atom has already
 a perfect coordination of 12, a further shortening of the distance
 between the BNN and the step atom causes the former to present a high frequency
 mode with  the highest amplitude. As mentioned, these conclusions
 about the high frequency modes have been verified  experimentally
 in the case of Cu(211) \cite{kar00}. While we are now engaged in
 performing {\it ab initio} calculations to further best the reliability
 of our results, we also await other experimental data
 to shed more light on the matter.

 In summary, a large reduction in the coordination of the step atoms
 yields a substantial softening of the force constants between step atoms
 along the same chain which causes a very strong and complex
 rearrangements of the atomic positions resulting in a shortening of
 the bond and hence also a stiffening of the force field near the step.
 These structural and dynamical changes localized in the vicinity
 of the step has important consequences for the thermodynamic
 properties of vicinal surfaces \cite{dur97}.
 Because
 of the localized nature of the characteristics presented here, we are now
 in a position to extend these conclusions to metallic nanoclusters 
 \cite{kar98}.

{\bf Acknowledgments}

This work  was  supported  by the Basic Energy Research
 Division
of the US Department of Energy.

\newpage

{\bf Figure Captions}

Figure 1: Geometry of Ni(977) :
side view (top)  and  top view
with surface Brillouin zone (bottom). SC= Step Chain, TC= Terrace Chain,
 CC= Corner Chain and BNN= Bulk Nearest Neighbor (of the step atom).

Figure 2: Local density of states for Ni(977) step and terrace atoms,
 and for a Ni(111) surface atom.

Figure 3 : Polarization of the Q1D mode at 3.3 THz .
The arrows pointing out of the spheres describe the amplitude of
the displacements. The moving atoms are transparent.

Figure 4 :  The z-component of the
 projected vibrational density of states, at $\Gamma$ (a),
$\overline{X}$ (b) and $\overline{Y}$ (c)  points for step
 (dashed),
 BNN (dot-dashed) and  bulk (solid line) atoms for Cu(211).

Figure 5: Shortening of the bond length between the step atom
 and it's neighbors for the three vicinals with terraces 3 atoms wide.

Figure 6: Shortening of the bond length between the step atom
 and it's neighbors for the three vicinals with terraces 4
 (Cu(711)), 5 (Cu(911)), and 9 (Cu(17,1,1)) atoms wide.

\newpage

\bigskip

\vskip .3in

\begin{table}
\caption{Illustration of the softening of the force field, 
 the force constants are in (eV/$\AA$/unit mass): Ni(977).}
\vskip .3in
\label{tab:forcon}
\begin{tabular}{lcrrrrrr}
\hline \hline
\multicolumn{1}{l}{ATOMS}&
\multicolumn{1}{l}{}&
\multicolumn{3}{c}{SURFACE}&
\multicolumn{3}{c}{BULK}\\
\hline
& &\multicolumn{1}{c}{$x$}&\multicolumn{1}{c}{$y$}&\multicolumn{1}{c}{$z$}&
\multicolumn{1}{c}{$\ \ \ \ \ \ x$}&\multicolumn{1}{c}{$y$}
&\multicolumn{1}{c}{$z$}\\
\hline
&$x$ & 0.0236&  0.1489&  0.0044&\ \ \ \ \ \ 0.0816&  0.0000&  0.0058\\
SC-SC&$y$&  -0.1489& \underline{-2.6386}&  -0.2796& 0.0000& \underline{-3.0568}&
 0.0000\\
&$z$ &0.0044&  0.2796&  -0.0050&  0.0058&  0.0000&  0.0805\\
\hline
\end{tabular}
\end{table}

\vskip 0.3in

\begin{table}
\caption{Calculated percentage changes ($\Delta$r) in bond length, its vertical
 component ($\Delta$z),
 between the step atom and its bulk nearest
 neighbor (BNN), for three Cu vicinal surfaces.  Here ($\Delta \nu$)
 is the shift in
 the frequency of the step localized mode above the maximum bulk mode.}
\vskip .1in
\begin{tabular}{cccc}
\hline \hline
  \multicolumn{1}{c}{Surface \hskip 46pt }  &
  \multicolumn{1}{c}{$\Delta r$ (\%) \hskip 29pt  }  &
  \multicolumn{1}{c}{$\Delta z$ (\%) \hskip 29pt }   &
  \multicolumn{1}{c}{$\Delta \nu$ (THz) \hskip 99pt } \\
\hline
 Cu(211) & -2.1  &  -2.7  &     + 0.24  \\
 Cu(511)   & -2.3  &  -2.9  &  + 0.24 \\
 \hskip 12pt Cu(17,1,1)   & -2.4  &  -2.7  &      + 0.43  \\
\hline
\end{tabular}
\end{table}

\vskip 0.3in

\begin{table}
\caption{Illustration of the stiffening of the force field, 
 the force constants are in (eV/$\AA$/unit mass): Cu(511).}
\begin{tabular}{lcrrrrrr}
\hline \hline
\multicolumn{1}{l}{ATOMS}&
\multicolumn{1}{c}{}&
\multicolumn{3}{c}{SURFACE}&
\multicolumn{3}{c}{BULK}\\
\hline
& &\multicolumn{1}{c}{$x$}&\multicolumn{1}{c}{$y$}&\multicolumn{1}{c}{$z$}&
\multicolumn{1}{c}{$\ \ \ \ \ \ x$}&\multicolumn{1}{c}{$y$}
&\multicolumn{1}{c}{$z$}\\
\hline
 &$x$&\underline{-2.1859}&0.0000&-0.8307&\ \ \ \ \ \ \underline{-1.8234}&0.0000&-0.5452\\
 SC-TC&$y$&0.0000&0.0678&0.0000&0.0000&0.0875&0.0000\\
 &$z$&-0.3896&0.0000&-0.1067&-0.5452&0.0000&-0.0499\\
\\
 &$x$&\underline{-0.9416}&-0.7116&0.6304&\underline{-0.8418}&-0.7037&0.7594\\
 CC-SC&$y$&-0.7637&-0.4282&0.4892&-0.7037&-0.4205&0.5598\\
 &$z$&0.8832&0.6337&-0.5432&-0.7594&0.5598&-0.5238\\
\\
 &$x$&-0.0683&0.3926&-0.5996&-0.0769&0.3062&-0.4846\\
 BNN-SC&$y$&0.3822&-0.5201&1.0726&0.3062&-0.4205&0.8454\\
 &$z$&0.6337&1.1648&\underline{-1.7441}&-0.4846&0.8454&\underline{-1.2888}\\
\hline
\end{tabular}
\end{table}

\begin{table}
\caption{Illustration of the stiffening of the force field,
 the force constants are in (eV/$\AA$/unit mass): Cu(211).}
\begin{tabular}{lcrrrrrr}
\hline \hline
\multicolumn{1}{l}{ATOMS}&
\multicolumn{1}{c}{}&
\multicolumn{3}{c}{SURFACE}&
\multicolumn{3}{c}{BULK}\\
\hline
& &\multicolumn{1}{c}{$x$}&\multicolumn{1}{c}{$y$}&\multicolumn{1}{c}{$z$}&
\multicolumn{1}{c}{$\ \ \ \ \ \ x$}&\multicolumn{1}{c}{$y$}
&\multicolumn{1}{c}{$z$}\\
\hline
 &$x$&\underline{-1.3793}&0.8954&-0.6340&\ \ \ \ \ \ \underline{-1.2838}&0.8500&-0.4907\\
SC-TC&$y$&0.8727&-0.4276&0.3859&0.8500&-0.4204&0.2932\\
 &$z$&-0.3259&0.2036&-0.1298&-0.4907&0.2932&-0.0818\\
\\
 &$x$&\underline{-1.6344}&0.0000&1.0355&\underline{-1.2837}&0.0000&0.9814\\
CC-SC&$y$&0.0000&0.0898&0.0000&0.0000&0.0875&0.0000\\
 &$z$&1.3683&0.0000&-0.7534&0.9814&0.0000&-0.5897\\
\\
 &$x$&0.1237&0.0162&-0.0582&0.0931&-0.0069&-0.0040\\
BNN-SC &$y$&-0.0140&-0.5070&1.1120&-0.0069&-0.4204&0.8991\\
 &$z$&-0.0770&1.9530&\underline{-1.9071}&-0.0040&0.8991&\underline{-1.4587}\\
\hline
\end{tabular}
\end{table}

\vskip 0.3in

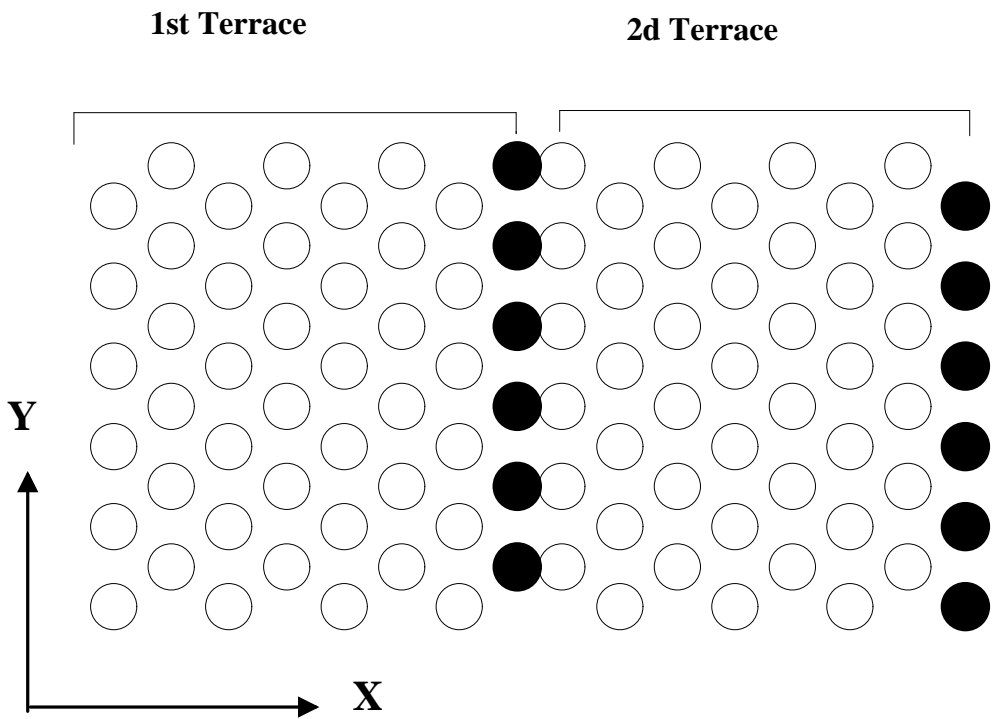
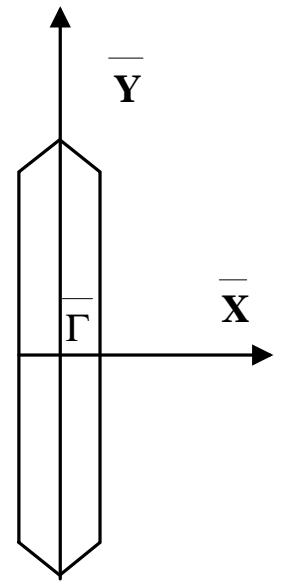
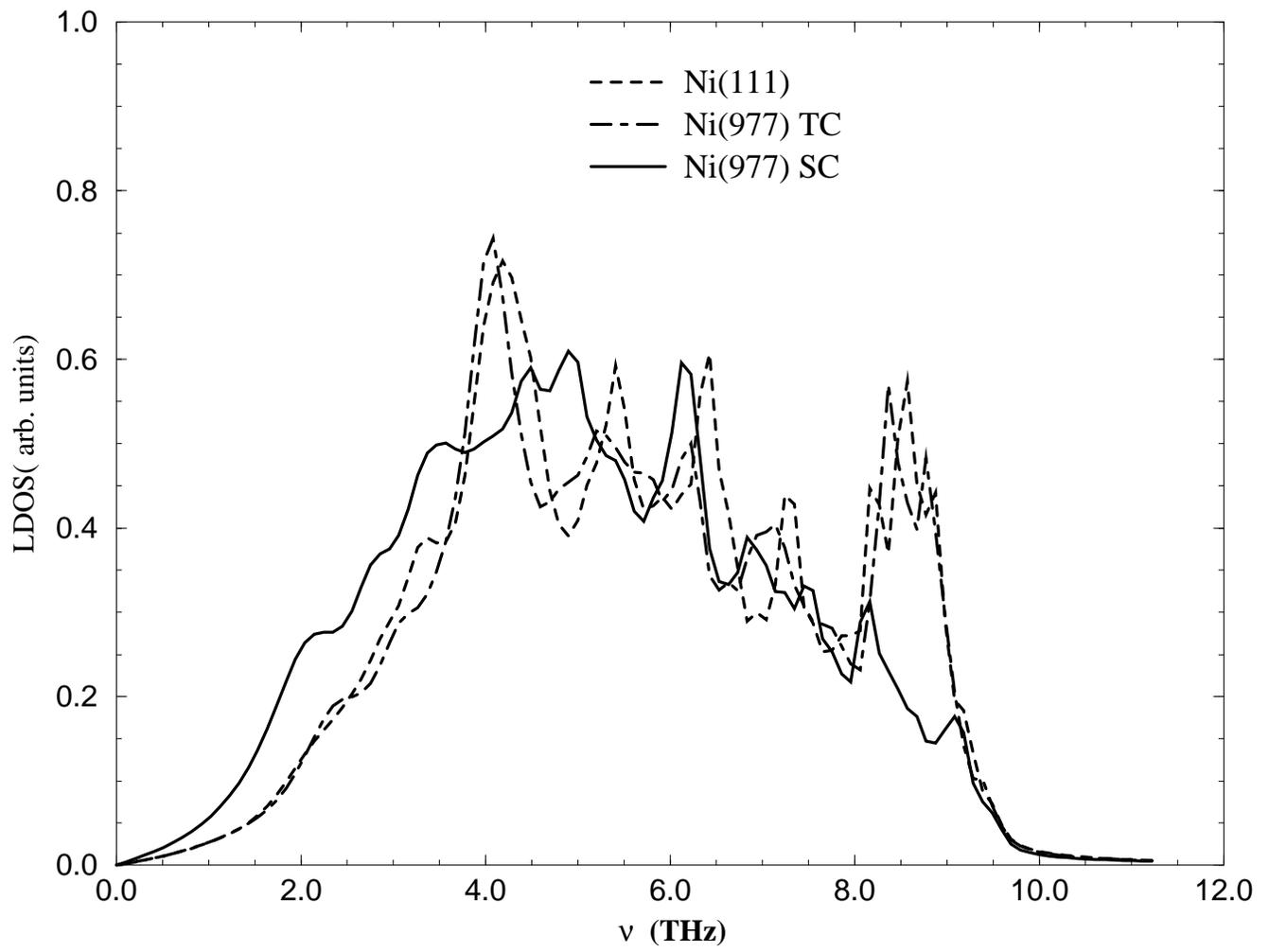
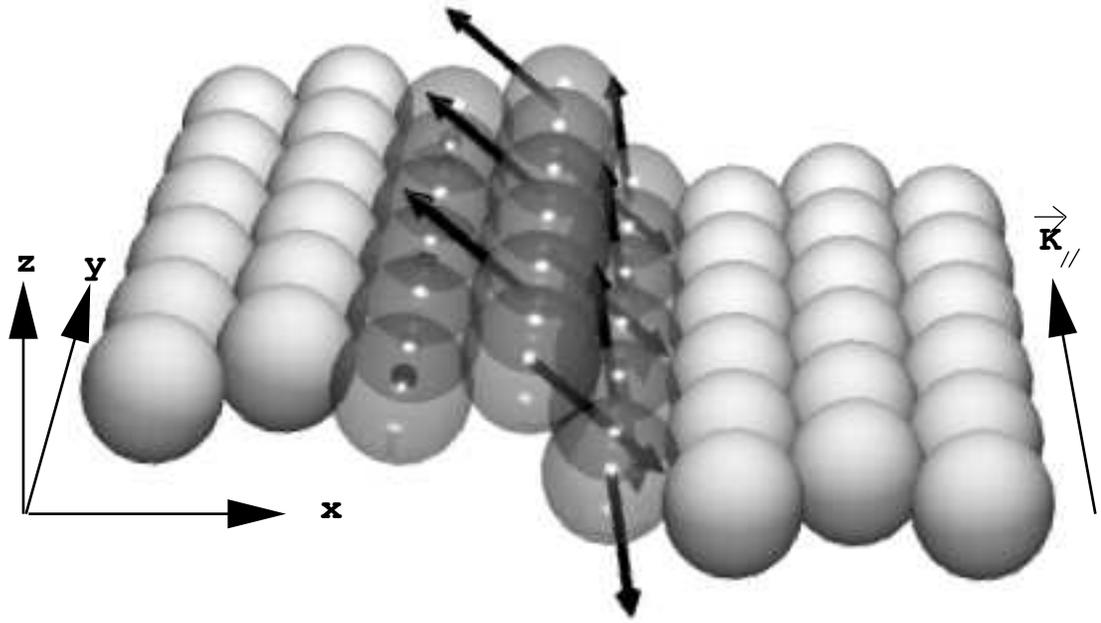
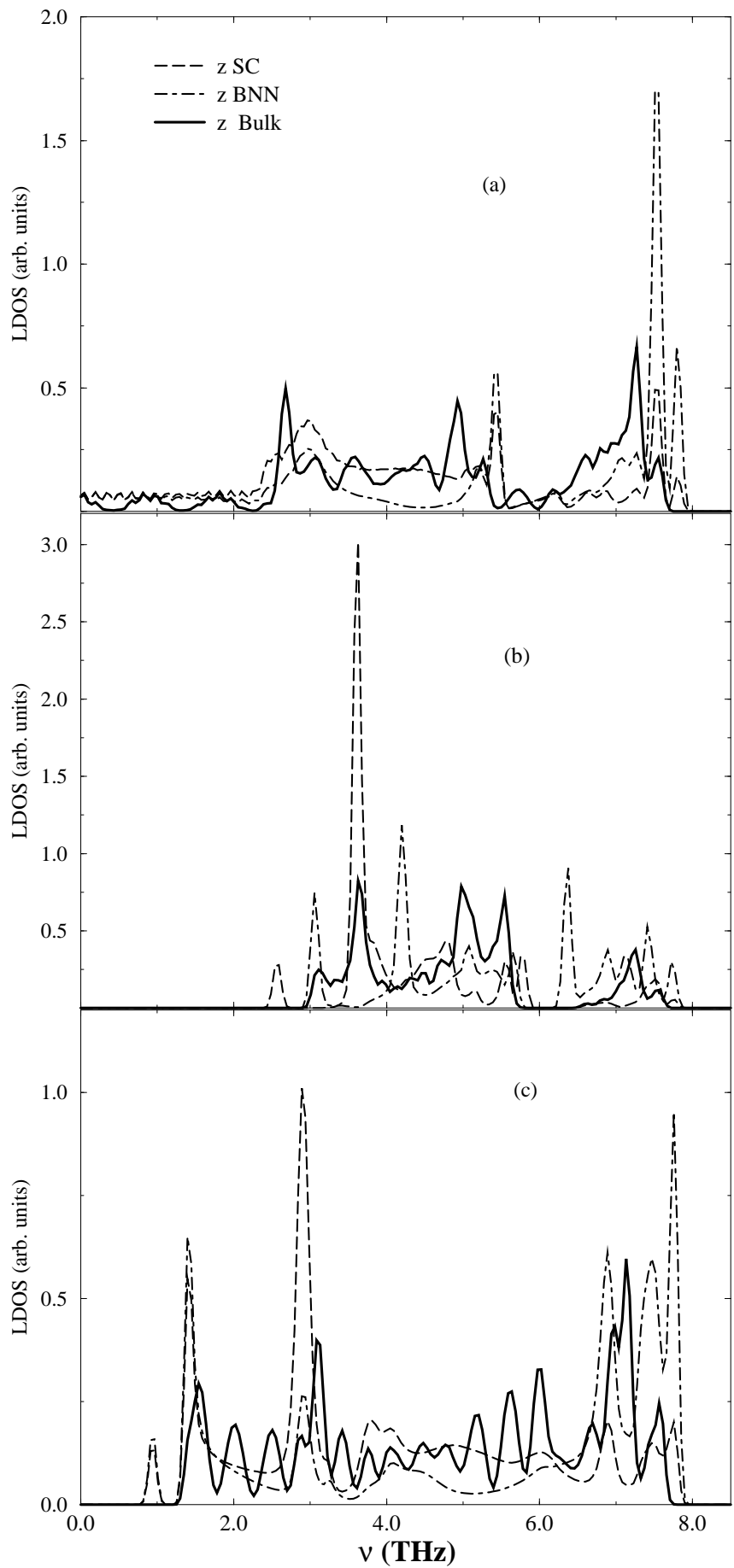
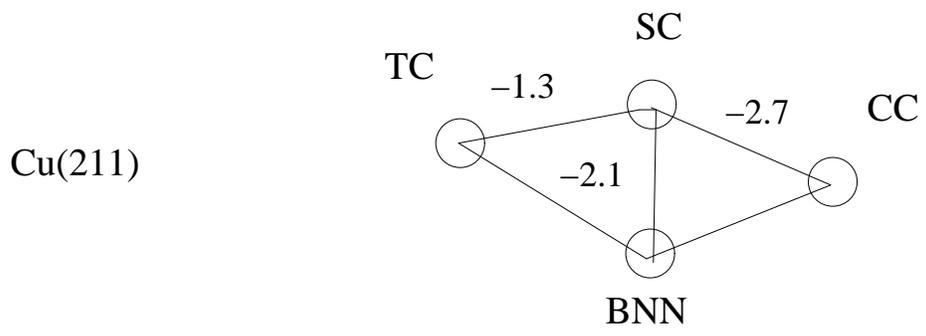
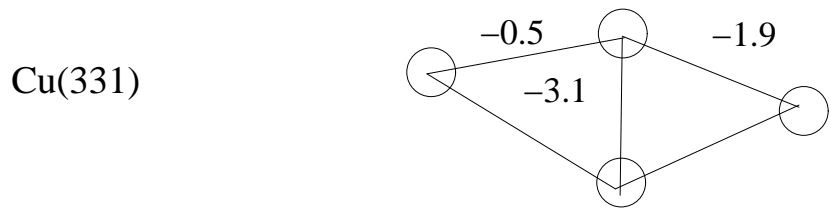
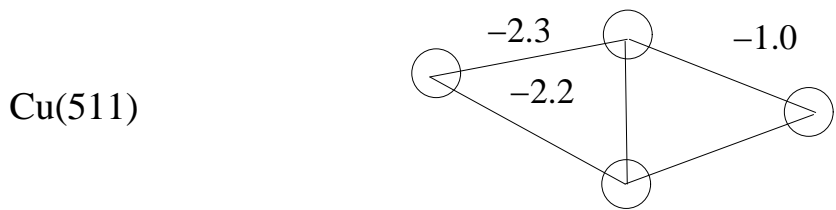
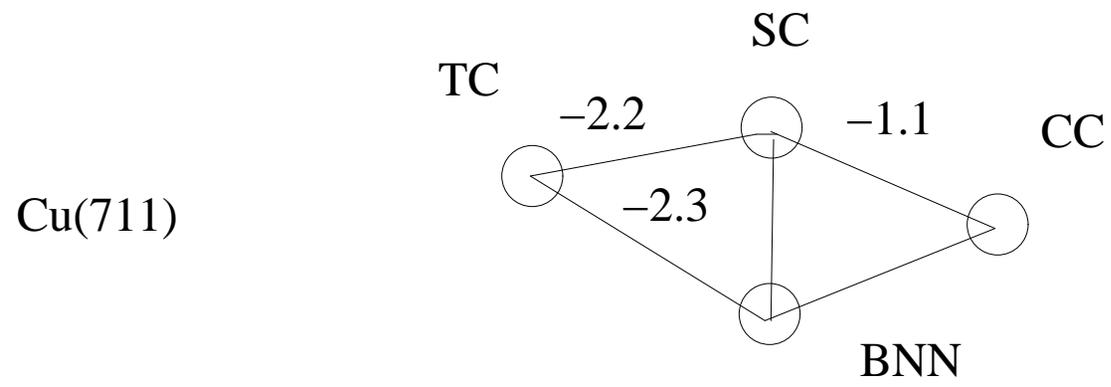
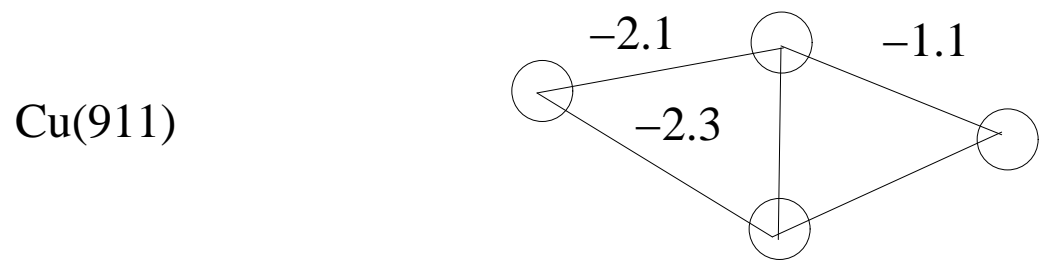
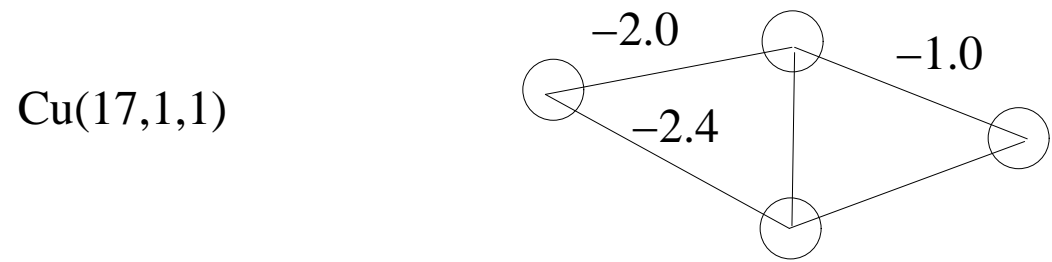
\end{document}